\documentstyle[eqsecnum,aps,prb,twocolumn,epsf]{revtex}

\begin{document}

\draft

\title{\hfill {\small Accepted for Publication in Phys. Rev. B} \\
  \vspace{10pt} Commensurability Effects in Large Josephson Junctions}
\author{Leon Balents}
\address{Institute for Theoretical Physics, University of California,
Santa Barbara, CA 93106-4030}
\author{Steven H.  Simon}
\address{Department of Physics, Harvard University, Cambridge, MA
  02138 \\ \vspace{.2in}\centerline{\begin{minipage}[t]{6.2in} \rm
 Two types of commensurability effects are possible in a large
 Josephson junction patterned with columnar defects.  The first occurs
 for a periodic array of pins when the mean fluxon spacing (tuned by
 the magnitude of the applied in--plane magnetic field) is a rational
 fraction of the defect spacing.  We show that this effect leads,
 under fairly general conditions, to a mapping of the behavior of the
 Josephson junction near the commensurate field values to that of a
 zero field junction with an effective Josephson penetration depth.
 The second occurs for more general arrangements of pinning sites,
 when the orientation of the Josephson vortex lattice (tuned by the
 direction of the applied field) nearly matches the orientation of the
 defects.  We investigate this tilt response in the limit of a single
 Josephson vortex.  The results are compared, where possible, to recent
 experiments.  As an aside from our main analysis, we prove that,
 contrary to recent claims in the literature, the critical current
 density vanishes in the thermodynamic limit, even in the presence of
 (non--pathologically distributed) pinning disorder. \end{minipage}}}
\maketitle

%\pacs{PACS: 74.60.Ge,74.40.+k}

\section{Introduction}
\label{intro}

The pursuit of large critical current densities in high temperature
superconductors in external magnetic fields has stimulated many
investigations into flux pinning.  Such pinning, caused by naturally
occurring and/or artificially induced non--superconducting defects, is
necessary to prevent dissipative flux flow in response to an applied
current.  At low temperatures in three dimensions, disorder is
expected to induce glassy phases, in which the current--voltage
relation is strongly non--linear, with vanishing linear resistivity.
These states for point and line pins have been named the vortex
glass\cite{VGtheory}\ and Bose glass,\cite{BGtheory}\ respectively.
Experimentally, continuous normal--glass transitions have been
observed for both types of disorder.\cite{VGexperiments,BGexperiments}

Quantitative comparison with the theory is, however, difficult for
several reasons.  On the theoretical side, neither the critical
exponents nor the properties of the low temperature phase are reliably
known.\cite{numerics}\ Experimentally, it is hard to prepare large
samples with well characterized disorder.  It is probably this
difficulty which has lead to a fairly broad range of measured values for
these critical exponents.

Josephson junctions (JJs) offer a promising alternative.  By orienting a
magnetic field parallel to the junction interface, the experimenter
can create Josephson vortices in the junction, which then acts in many
ways like a 1+1 dimensional version of the bulk vortex state.  Because
of the sophistication and ease of modern fabrication techniques,
samples may be engineered with almost any configuration of defects in
an extremely controlled way.  The theoretical situation is also much
more secure for 1+1 dimensional glasses, mainly due to the ability to
perform renormalization group calculations.\cite{RGrefs}\ Large scale
numerical simulation in 1+1 dimensions are also more tractable than
their 2+1 dimensional counterparts.\cite{HwaBatrouni}\

Patterned JJs are interesting in their own right, and have been
studied by several authors.\cite{FGB,VK}\  Here, we discuss the
case of linear (columnar) defects, focusing on commensurability
effects tuned by changes in the magnetic field.  Because of the
relative ease of controlling the magnetic field versus other system
parameters (e.g. temperature, disorder), such effects provide a
convenient probe of the properties of the JJ.

The remainder of the paper is organized as follows.  In section
\ref{JJreview}, we review the properties of large uniform JJs and the
equations describing their behavior.  Section \ref{Defects}\ describes
how defects are incorporated into this model.  In addition, we prove
that, other than for particularly pathological cases, the introduction
of defects {\sl does not} allow a non--zero average critical current
{\sl density} for the JJ in the thermodynamic limit, contrary to
recent claims in the literature.\cite{FGB,VK}\ Commensurability
effects which occur when varying the magnitude of the applied magnetic
field in a periodic defect array are explained in section
\ref{PeriodicDefects}.  Extensive experiments on such systems have
been carried out in Refs.\onlinecite{Itzler}\ and
\onlinecite{Itzler2}, in good agreement with the theoretical results
presented here.  A discussion of complex commensurate states which
occur at 1/2--integral fields is included in Appendix A.

A qualitatively different type of commensurability effect occurs upon
changing the direction of the magnetic field.  In section
\ref{TiltResponse}, we calculate this tilt response for an
isolated flux line, where the well--known analogy\cite{Seung}\ to a
one dimensional system is particularly useful.  In section
\ref{Conclusion}, we conclude with a discussion of the experimental
checks on this work and its ramifications for future research.

\section{Large Josephson Junctions}
\label{JJreview}

A Josephson junction is an insulating interface between two (usually
identical) superconductors.  For simplicity, we will restrict the
discussion to the so--called {\sl overlap} geometry, shown in
Fig.1.  This is a common experimental choice, but by no
means unique.  Other configurations behave similarly, but require
different boundary conditions.

%\begin{figure}
%\epsfxsize=3.5truein
%\hskip 0.0truein \epsffile{geom.eps}
% { FIG 1: Overlap geometry for a large Josephson junction.}
%\label{overlapfig}
%\end{figure}

In a JJ, all the quantities of interest may be expressed in terms of
the gauge--invariant phase difference $\gamma \equiv \theta_1 -
\theta_2 - 2\pi/\phi_0 \int_{-d/2}^{d/2} dz A_z(z)$ between the two
superconductors, where $\phi_0 = hc/2e$ is the flux
quantum.\cite{Tinkham}\ Here $\theta_1$ and $\theta_2$ are the phases
of the BCS order parameters in the top and bottom superconducting
slab, $A_z$ is the z--component of the vector potential. The length
$d$ is the magnetic thickness of the junction, and is related to the
actual thickness $s$ by $d = s + 2\lambda$, where $\lambda$ is the
London penetration depth.  The current density in the $z$--direction
has the Josephson form
\begin{equation}
j_z = j_0 \sin \gamma.
\label{josephsoncurrent}
\end{equation}
We will regard the tunneling current density $j_0$ as an experimental
constant.  The magnetic field components parallel to the junction
plane are
\begin{eqnarray}
B_x & = - & {\phi_0 \over {2\pi d}} {{\partial \gamma} \over
{\partial y}} \nonumber \\
B_y & = & {\phi_0 \over {2\pi d}} {{\partial \gamma} \over
{\partial x}},
\label{magneticfield}
\end{eqnarray}
Assuming a static situation in the
interior of the junction, Eqs.\ref{josephsoncurrent}\ and
\ref{magneticfield}\ can be combined with the Maxwell equation $\bbox{\nabla}
\times {\bf B} = 4\pi {\bf j}/c$ to yield
\begin{equation}
\lambda_J^2 \nabla^2 \gamma = \sin\gamma,
\label{sinegordon}
\end{equation}
where $\nabla^2$ indicates the two dimensional Laplacian in the
interface plane.  The length $\lambda_J = \sqrt{c\phi_0/8\pi^2dj_0}$
is known as the Josephson penetration depth.  Eq.\ref{sinegordon}\ is
known as the (time independent) sine--Gordon equation.

Assuming the current is injected parallel to the $y$--axis, the
boundary conditions augmenting Eq.\ref{sinegordon}\ are\cite{OS}
\begin{eqnarray}
\left.{{\partial\gamma} \over {\partial x}}\right|_{x=0} & = & {{2\pi
d H_y} \over \phi_0} \nonumber \\
\left.{{\partial\gamma} \over {\partial x}}\right|_{x=W} & = & {{2\pi
dH_y}  \over \phi_0} \nonumber \\
\left.{{\partial\gamma} \over {\partial y}}\right|_{y=0} & = & -{{2\pi
dH_x} \over \phi_0} - {{4\pi^2dI} \over {c\phi_0W}}
\nonumber \\
\left.{{\partial\gamma} \over {\partial y}}\right|_{y=L} & = & -{{2\pi
dH_x} \over \phi_0} + {{4\pi^2dI} \over {c\phi_0W}}
\label{bcs}
\end{eqnarray}

Eq.\ref{sinegordon}, in conjunction with Eq.\ref{bcs}, can have
multiple solutions.  In such cases, the equilibrium situation is found
by minimizing the free energy
\begin{equation}
F = \epsilon_{\rm J}\int dx dy \left\{ {1 \over 2}|\bbox{\nabla}\gamma
- {\bf h}|^2 +
\lambda_J^{-2}\left( 1 - \cos\gamma \right) \right\},
\label{freeenergy}
\end{equation}
where $\epsilon_J = \phi_0^2/16\pi^3d$ is the overall energy scale and
${\bf h} = 2\pi d(-H_y,H_x)/\phi_0$.

We next briefly summarize the properties of a defect free
junction.\cite{JJgeneral}
In the absence of an applied magnetic field, a {\sl small} or {\sl
long} junction with $L \lesssim \lambda_J$ can carry a current $I =
LWj_0\sin\gamma$, with $\gamma$ approximately uniform, leading to a
critical current $I_{c0} = LWj_0$.  This can be seen by transforming
away the boundary conditions according to $\gamma \rightarrow \gamma +
I(y-L/2)^2/(2j_0LW\lambda_J^2)$.  For {\sl large} junctions, where
{\sl both} $W,L \gg \lambda_J$, the uniform approximation fails.  In
this limit, the phase (and therefore the field and current) decays
exponentially to zero in the interior of the junction.  This screening
is a property of the linearized sine--Gordon equation,
\begin{equation}
\lambda_J^2 \nabla^2 \gamma \approx \gamma.
\label{linearizedSG}
\end{equation}
In this regime, $I_c \propto \lambda_J^2 j_0$.

In an applied magnetic field, the critical current is further reduced.
For strong fields, $H \gg \phi_0/d\lambda_J$, screening is negligible,
and the phase winds approximately linearly transverse to the field
axis, e.g.
\begin{equation}
\gamma \approx \hat{\gamma} + {{2\pi d} \over \phi_0}H_y x, \; \; {\rm for}
\;\; H_x=0, H_y \gg \phi_0/d\lambda_J.
\label{strongfield}
\end{equation}
In this limit, the critical current, obtained most simply by
integrating Eq.\ref{josephsoncurrent}, takes the Fraunhofer form
\begin{equation}
I_c \approx j_0 LW\left| {{\sin(\pi d H_y W/\phi_0)} \over {\pi d H_y
W/\phi_0}} \right|.
\label{fraunhofer}
\end{equation}
For $H \gtrsim \phi_0/d\lambda_J$, the field penetrates the junction,
but not uniformly.  Instead, it gathers into ``Josephson vortices'' or
``solitons'' of total flux $\phi_0$, across which $\gamma$ changes by
$2\pi$.  For large fields, these vortices comprise only a weak
sinusoidal modulation of the field in the junction (as can be obtained
by perturbation theory from Eq.\ref{sinegordon}).

As $H \rightarrow H_{c1J}$, with $H_{c1J} = 2 \phi_0/\pi^2 d\lambda_J$,
however, the solitons sharpen into objects of definite width
$\lambda_J$, between which $\gamma$ rapidly decays to a multiple of
$2\pi$.  For $(H-H_{c1J})/H_{c1J} \ll 1$, a kind of critical
phenomenon, known as a commensurate--incommensurate transition (CIT),
governs the vanishing density of Josephson vortices.  In this regime,
it is legitimate to treat the solitons as weakly interacting, since
their average separation is much larger than $\lambda_J$.  The scaling
of the free energy of an {\sl isolated} Josephson vortex can be read
off from Eq.\ref{freeenergy}.  By using the exact one--soliton
solution to Eq.\ref{sinegordon},
\begin{equation}
\gamma_0(x) = 4\tan^{-1}(\exp(x/\lambda_J)),
\label{exactonesoliton}
\end{equation}
and subtracting the free energy for a configuration with no vortex,
the full expression including the prefactor is found as
\begin{equation}
F_{vortex} = -8 (H/H_{c1J}-1){L \over \lambda_J}\epsilon_J.
\label{isolatedfreeenergy}
\end{equation}
For $H>H_{c1J}$, of course, inter--soliton interactions must be
included to prevent the vortices from proliferating.  In addition, we
will also allow for small fluctuations in the positions and shapes of
the vortices, since sufficiently small deformations of $\gamma$ cost a
free energy less than $k_{\rm B}T$.  Doing so leads to (neglecting an
unimportant constant term) the soliton free energy (for $N$ vortices)
\begin{eqnarray}
{\cal F}_N & = & -N(H/H_{c1J}-1)\tilde{\epsilon} L + \int_0^L dy \bigg\{
\sum_{n} {\tilde{\epsilon} \over 2}\left|{{dx_n(y)} \over
{dy}}\right|^2
\nonumber \\
& & +
\tilde{\epsilon}\sum_{n,n'} V(|x_n(y)-x_{n'}(y)|/\lambda_J) \bigg\},
\label{vortexfreeenergy}
\end{eqnarray}
where $\tilde{\epsilon} \equiv 8\epsilon_J/\lambda_J$, and $V(\chi)
\sim 2\chi\exp(-\chi)$ is
an exponentially decaying repulsive interaction with magnitude and
range of order unity (we have approximated the interactions as local
in $y$, which is adequate for the dilute limit considered
here).\cite{lineenergynote}\

To understand the vanishing of the vortex density at the CIT,
Eq.\ref{vortexfreeenergy}\ is used to estimate the free energy as a
function of $\ell$, the average fluxon separation.  At finite
temperature, an entropic contribution must be included due to the
wandering of the solitons along the $y$ axis.\cite{CIT} By
equipartition, an individual Josephson vortex wanders according to
$\langle (x(y)-x(0))^2\rangle \sim k_{\rm B}T y/\tilde{\epsilon}$.
Each time the fluxon wanders a transverse distance $\ell$, it is caged
by another soliton, and must reverse direction Naively, this
constraint excludes half of the available configurations of the
vortex.  Taking the logarithm of the total number of configurations
gives the entropy per vortex, which therefore is reduced by a constant
$\sim k_{\rm B}\ln 2$.  This leads to a free energy cost $-T \Delta S
\sim (k_{\rm B}T)^2 L/(\tilde{\epsilon}\ell^2)$ for a single Josephson
vortex.  Including this with the energetic contributions from the
first and last terms in Eq.\ref{vortexfreeenergy}\ gives the free
energy density (per unit area $LW$)
\begin{equation}
f \sim -r{\epsilon_J \over {\ell\lambda_J}} + {\epsilon_J \over
\lambda_J^2}e^{-\ell/\lambda_J} + {{(k_{\rm B}T)^2} \over
\epsilon_J}{\lambda_J \over \ell^3},
\label{CITfreeenergy}
\end{equation}
where $r=(H/H_{c1J}-1)$.  For $k_{\rm B}T \ll \epsilon_J$, the minimum
free energy is determined by the balance between the first and second
terms, leading to
\begin{equation}
\ell_{T=0} \sim \lambda_J\ln(1/r),
\label{ellgrowth}
\end{equation}
or, in terms of the internal average magnetic field, $\langle B
\rangle \sim \phi_0/(\lambda_J d \ln(1/r))$.  For sufficiently small
$r$, or at high temperatures, however, this law breaks down, due to
the effects of the entropic term.  The true asymptotic behavior is
obtained by balancing the first and last terms to give
\begin{equation}
\ell_{\rm asymp.} \sim {{k_{\rm B}T} \over \epsilon_J} {\lambda_J
\over \sqrt{r}}.
\label{asymptoticellgrowth}
\end{equation}
The crossover to this behavior is, however, extremely close to the CIT
at typical experimental temperatures.  It occurs when $r < r_{co} \sim
(k_{\rm B}T/\epsilon_J)^2/\ln^3(\epsilon_J/k_{\rm B}T)$, which for,
say, a niobium junction with $d \approx 1500 \AA$ at $T=4.2K$,
requires $r \lesssim 10^{-8}$.  These thermal effects might, however,
be more observable near the bulk superconductor--normal transition
(especially in high temperature superconductors) where the Josephson
coupling energy vanishes (within mean field theory) like
$\sqrt{1-T/T_c}$.

\section{Junctions with Defects}
\label{Defects}

The junctions of Ref.\onlinecite{Itzler}\ contain defects, which are
regions in which the local thickness is substantially increased.
For columnar defects parallel to the $y$ axis, Eq.\ref{josephsoncurrent}\
must be modified to $j_0 \rightarrow \alpha(x)j_0$, with $\alpha(x)
\leq 1$.  We have assumed that the patterning is uniform in the $y$
direction (columnar defects), so that $\alpha$ is independent of $y$.
The tunneling current has a strong (exponential) dependence when the
local thickness variation $\delta d \gtrsim \xi$, the superconducting
coherence length.  The expressions for the magnetic field
(Eq.\ref{magneticfield}) also become position dependent through the
factor of $1/d(x)$, but we will ignore this weaker dependence in what
follows.\cite{thicknessnote}

The modulation in $j_0(x)$ results in a modified sine--Gordon
equation,
\begin{equation}
\lambda_J^2\nabla^2 \gamma = \alpha(x)\sin\gamma.
\label{modifiedSG}
\end{equation}
The boundary conditions, Eq.\ref{bcs}, are unchanged.

The modified sine--Gordon equation leads naturally to defect--related
corrections to the dilute fluxon free energy,
Eq.\ref{vortexfreeenergy}.  To compute them, we first write the
modified sine--Gordon free energy corresponding to
Eq.\ref{modifiedSG},
\begin{equation}
F = \epsilon_{\rm J}\int dx dy \left\{ {1 \over 2}|\bbox{\nabla}\gamma
- {\bf h}|^2 +
\lambda_J^{-2}\alpha(x)\left( 1 - \cos\gamma \right) \right\}.
\label{modifiedSGFE}
\end{equation}
Inserting $\gamma(x,y) = \sum_i\gamma_0(x_i(y))$, keeping only the
leading (one--body) term and ignoring an additive constant, gives a
correction to Eq.\ref{vortexfreeenergy}\ of
\begin{equation}
\Delta {\cal F}_N = \tilde{\epsilon}\int dy \sum_n U(x_n(y)),
\label{defectfreeenergy}
\end{equation}
where
\begin{equation}
U(x) \equiv {1 \over 4}\int {{dx'} \over \lambda_J} {{\alpha(x+x')} \over
{\cosh^2(x'/\lambda_J)}}.
\label{effectivepotential}
\end{equation}

It has been pointed out by several authors\cite{FGB,VK} that an
appropriate spatially dependent $\alpha(x)$ can lead to an {\sl
increase} in the critical current of a JJ.  In section
\ref{PeriodicDefects}\ we will demonstrate this explicitly for the
case of periodic defects.  In the remainder of this section, however,
we will demonstrate that, contrary to several claims\cite{FGB,VK}, the
critical current density of a {\sl large} ($W,L \gg \lambda_J$) JJ
with columnar defects is always zero in the thermodynamic
limit.\cite{pathology} The critical current density also vanishes for
the {\sl long} ($W \gg
\lambda_J$, $L \ll
\lambda_J$) ``in--line asymmetrical'' geometry considered in
Ref.\onlinecite{FGB}.\cite{longnote} The vanishing of the critical
current density results from the screening of current inherent in the
sine--Gordon equation.  Even in zero applied magnetic field in a {\sl
pure} junction, the critical current density decreases with the system
size once it is larger than $\lambda_J$.

To prove that the critical current density indeed vanishes, we will
take a more concrete model of the disorder.  Specifically, $\alpha(x)$
is taken to be piecewise constant on intervals of variable length.  On
all intervals, we require $0< \alpha < 1$, the upper bound
corresponding to a perfect junction (in fact, it is only necessary
that $\alpha$ is bounded above and below by arbitrary constants).  The
lengths of the intervals should also be well behaved, so that the mean
length $\sum_i x_i/N = \bar{x}$ exists.  For simplicity, we consider the
one--dimensional junction (i.e. $L \ll \lambda_J$) studied in
Ref.\onlinecite{FGB}.

Consider the mean current density for the junction,
\begin{equation}
\bar{j} \equiv {1 \over W}\int_0^W dx j_0 \alpha(x)
\sin\gamma,
\label{meancurrentdensity}
\end{equation}
where $\gamma$ is the solution of the one--dimensional sine--Gordon
equation,
\begin{equation}
\lambda_J^2 {{d^2\gamma} \over {dx^2}} = \alpha(x)\sin\gamma.
\label{onedSG}
\end{equation}
Multiplying by $j_0$ and integrating over $x$ gives
\begin{equation}
\bar{j} = j_0\lambda_J^2(v(W) - v(0))/W,
\label{jbarequ}
\end{equation}
where $v(x) = d\gamma(x)/dx$.  Multiplying Eq.\ref{onedSG}\ instead by
$\lambda_J^{-2}v(x)$ and integrating from $x$ to $x'$, one finds
\begin{equation}
E(x') - E(x) = \lambda_J^{-2}\sum_{x<x_i<x'} \delta\alpha_i \cos\gamma(x_i),
\label{firstintegral}
\end{equation}
where $x_i$ is the coordinate of the boundary between the $i^{th}$ and
$(i+1)^{th}$ constant region, and $\delta\alpha_i$ is the jump in
$\alpha$ at that boundary.  The function $E(x) = v(x)^2/2 +
\alpha(x)\cos\gamma(x)$.  Using $|\delta\alpha_i| \leq 1$ and the
existence of $\bar{x}$, the magnitude of the right hand side of
Eq.\ref{firstintegral}\ is bounded above by $\lambda_J^{-2}N =
\lambda_J^{-2}|x-x'|/\bar{x}$ (for large $L$).  Since the $\cos\gamma$
term in $E(x)$ is order one, we have
\begin{equation}
|v(x')^2 - v(x)^2| \leq C\lambda_J^{-2}|x-x'|/\bar{x}.
\label{v2bound}
\end{equation}
Here and in the remainder of this section, $C$ indicates any
dimensionless constant of order one.  Consider first the case where
$v(W)$ and $v(0)$ have the same sign.  Then $|v(W)^2-v(0)^2| =
|(v(W)-v(0))(v(W)+v(0))| > |v(W)-v(0)|^2$.  Using Eq.\ref{jbarequ},
this gives the bound $|\bar{j}| \leq C j_0\lambda_J/\sqrt{\bar{x}W}$.
If $v(W)$ and $v(0)$ have opposite signs, then there exists some
$x^*$ between $0$ and $W$ with $v(x^*)=0$.  Both $x^*<W$ and
$W-x^*<W$, so $|v(W)-v(0)| = |v(W)| + |v(0)| \leq
2C\lambda_J^{-2}W/\bar{x}$, which leads to a similar bound.  We have
therefore established in general that
\begin{equation}
|\bar{j}| \leq C j_0 {\lambda_J \over \sqrt{\bar{x}W}},
\label{jbarlimit}
\end{equation}
which vanishes in the thermodynamic ($W \rightarrow \infty$) limit.

This result can be understood in terms of an amusing analogy.
Consider a simple pendulum in a time varying gravitational field.  If
the gravitational constant cannot become negative or exceed it's
normal value, Eq.\ref{onedSG}\ describes this situation with $\gamma$
giving the angle of the pendulum and $x$ taking the role of time.  By
Eq.\ref{jbarequ}, a non--zero critical current density would be
equivalent to saying that one could continuously accelerate the
pendulum by varying the gravitational field.  Clearly, as the pendulum
begins to move faster, the field must be switched from strong to weak
more and more rapidly to continue to give it an average acceleration.
This would require a vanishing $\bar{x}$, corresponding to the
infinitesimal time interval in the large $W$ limit.

\section{Periodic Defects}
\label{PeriodicDefects}

Although a finite critical current density is impossible for an
infinite system, an increased critical current can be obtained in
finite systems by introducing pinning.  A natural choice for such
pinning is to choose a defect periodicity commensurate with the
Josephson vortex spacing in the junction.  Such junctions have been
studied extensively in Refs.\onlinecite{Itzler,Itzler2},
and we discuss a theoretical approach to these junctions here.

For ease of presentation, we will assume that the defect configuration
is such that the unit cell can be chosen even.  The extension to unit
cells without symmetry is straightforward.  In this case, the phase is
governed by Eq.\ref{modifiedSG}, where $\alpha(x)$ can be written in
the form
\begin{equation}
\alpha(x) = \sum_{m=0}^\infty \alpha_m \cos mqx,
\label{alphaexpand}
\end{equation}
where $q=2\pi/a$, with defect lattice spacing $a$.  Making the change
of variables $\gamma = nqx + \eta$ and inserting
Eq.\ref{alphaexpand}, Eq.\ref{modifiedSG} becomes\cite{integralstatenote}\
\begin{eqnarray}
\lambda_J^2 \nabla^2 \eta & = & \sum_{m=0}^\infty {\alpha_m \over 2}
\bigg\{ \sin\left[ (n+m)qx + \eta \right] \nonumber \\
& & + \sin\left[ (n-m)qx + \eta \right] \bigg\}.
\label{expandedSG}
\end{eqnarray}
Under this transformation, the form of the boundary conditions
(Eq.\ref{bcs}) remains unchanged (with $\eta$ replacing $\gamma$),
except that the $y$ component of the magnetic field is replaced by an
effective value
\begin{equation}
\tilde{H}_y = H_y - {{nq\phi_0} \over {2\pi d}}.
\label{effectivefield}
\end{equation}
To simplify further, we will assume that the defect lattice is not too
large, so $a \lesssim \lambda_J$, and that the field has been chosen
close to commensurate, so that $\tilde{H}_y d\lambda_J \ll \phi_0$.
Under these assumptions, $\eta$ will be slowly varying, and all the
sines but the second $m=n$ term will average
out\cite{correctionsnote}\ (unless $n=0$, in which case both $m=0$
terms are equal).  Eq.\ref{expandedSG}\ then reduces to a uniform
sine--Gordon equation,
\begin{equation}
\tilde{\lambda}_J^2 \nabla^2 \eta = \sin\eta,
\label{effectiveSG}
\end{equation}
where the effective Josephson penetration depth is
\begin{equation}
\tilde{\lambda}_J = \cases{ \sqrt{2/|\alpha_n|}\lambda_J & $n \neq 0$
\cr \sqrt{1/\alpha_0}\lambda_J & $n=0$}.
\label{effectivelambdaJ}
\end{equation}
Although we have not discussed time dependence in any detail, exactly
the same mapping holds if it is included through the usual
$\lambda_J^2(\bar{c}\partial_t^2 + \beta\partial_t)\gamma$ terms on
the left hand side of Eq.\ref{sinegordon}.  Thus the statics {\sl and}
dynamics near the commensurate fields will behave as in a uniform
junction, with an $n$--dependent effective penetration depth and
shifted magnetic field.  For small $\tilde{H}_y$, the physics will
therefore extremely closely mimic that of the uniform case.  In
particular, the critical current at the $n^{th}$ peak is simply
evaluated at zero field.  Thus
\begin{equation}
I_{cn} = f(L/\tilde{\lambda}_J)LW\tilde{j_0} =
f(L/\tilde{\lambda}_J)LWj_0{\lambda_J^2 \over \tilde{\lambda}_J^2},
\label{commensuratecritcurr}
\end{equation}
where $f(\chi) \rightarrow 1$ for $\chi \lesssim 1$ and $f(\chi)
\sim 2/\chi$ for $\chi \gg 1$.

As an example, we explicitly compute the effective parameters for the
case studied experimentally in Ref.\onlinecite{Itzler}.  There, the
thickness of the defects is large enough ($d \gg \xi$) to render
$\alpha=0$ on the pinning sites.  In this case, the function
$\alpha(x)$ is a periodic train of rectangular pulses of width $a-w_d$
and height $1$, where $w_d$ is the defect width, separated by spaces
of width $w_d$ with $\alpha=0$.  Choosing the origin at the center of
a defect, the Fourier series in Eq.\ref{alphaexpand}\ can be inverted
to yield
\begin{equation}
\alpha_n = \cases{-2\sin(\pi n w_d/a)/(\pi n), & $n>0$ \cr
1-w_d/a, & $n=0$}.
\label{itzlercase}
\end{equation}
This implies an effective Josephson penetration depth of
\begin{equation}
\tilde{\lambda}_J = \cases{\lambda_J\sqrt{\pi n/\sin(\pi n
w_d/a)}, & $n>0$ \cr
\lambda_J/\sqrt{1-w_d/a}, & $n=0$}.
\label{lambdaeffectiveitzler}
\end{equation}
For an effectively long or small junction (note that this only
requires $\tilde{\lambda}_J \gtrsim L$, {\sl not} $\lambda_J \gtrsim
L$), this gives the critical current $I_c = LWj_0|\sin(\pi n
w_d/a)/(\pi n)|$ for $n>0$ and $I_c = LWj_0(1-w_d/a)$ for $n=0$.  In
this limit, these results have been obtained in
Ref.\onlinecite{Itzler}\ using a simple current blocking model.  We
point out here that the region of validity of these results is
actually much wider than would be naively predicted, because the
increased effective Josephson penetration depth brings the system
further into the small junction limit.  The predicted modification of
the commensurate critical current due to the increased effective
screening length (obtained from Eqs.\ref{commensuratecritcurr}\ and
\ref{lambdaeffectiveitzler}) has recently received strong experimental
support.\cite{Itzler2}\

Further, we can discuss in this way the width of the commensurate
peaks.  For $\tilde{\lambda}_J \ll W$, in the overlap geometry
considered here, the width of a peak is determined by the screening
condition $\tilde{H}_y \sim \tilde{H}_{c1J} \sim
\phi_0/d\tilde{\lambda}_J$.  Using Eq.\ref{lambdaeffectiveitzler},
this gives
\begin{equation}
\Delta H_{y,n}^{\rm screened} \sim \cases{{\phi_0 \over
{d\lambda_J}}\sqrt{\sin(\pi n
w_d/a)/n\pi}, & $n>0$ \cr
{\phi_0 \over {d\lambda_J}}\sqrt{1-w_d/a}, & $n=0$}.
\label{peakwidthsinscreenedlimit}
\end{equation}
We note in passing that an in--line geometry, in which the current is
injected transverse to the magnetic field, has considerably different
behavior in this limit, due to the restriction of the current to
screening layers near $x=0,W$.  In the unscreened limit (still the
overlap geometry) where $\tilde{\lambda}_J \gg W$,
\begin{equation}
\Delta H_{y,n}^{\rm unscreened} \sim {\phi_0 \over {dW}} .
\label{peakwidthsinunscreenedlimit}
\end{equation}
Between the two scaling regimes there is presumably a smooth crossover
(which could, in principle, be calculated from a solution of the
uniform sine--Gordon equation).

\section{Tilt Response of a Single Flux Line}
\label{TiltResponse}

As discussed in section \ref{JJreview}, the fluxon density vanishes
continuously as $H$ is reduced toward $H_{c1J}$.  Once the mean
spacing $\ell \gtrsim 10\lambda_J$, say, the exponentially decaying
interactions between Josephson vortices are weak, and it is
reasonable to attempt to approximate the solitons as independent.  In
fact, at finite temperature in an infinitely long ($L=\infty$) system,
thermally excited transverse wandering of the vortices actually
invalidates this approximation.  In normal experimental conditions
away from the bulk superconductor--normal transition, however, $k_{\rm
B}T/\epsilon_J$ is so small (c.f. the paragraph after
Eq.\ref{asymptoticellgrowth}) that this fluctuation effect is
negligible for  typical sample lengths.  A more serious problem is the
extremely weak divergence of $\ell$ near $H_{c1J}$ given in
Eq.\ref{ellgrowth}\ means that it is necessary to reach a reduced
magnetic field of  $r \lesssim e^{-10} \approx 5 \times 10^{-5}$ to
get to this regime.

It is nevertheless interesting from a theoretical point of view to
consider the limit of an isolated flux line.  Such a situation is
somewhat more realizable in three dimensions in a bulk superconductor,
a case into which we may hope to gain insight through this simpler
model.  It also represents a soluble limit, even in the case of
a random distribution of defects.  We discuss this interesting random
case next at $T=0$.

For simplicity, we define the slope of the applied magnetic field $h \equiv
\frac{H_x}{H_y}$.  Combining Eq.\ref{vortexfreeenergy}\ and
Eq.\ref{defectfreeenergy}, the free energy for a single flux line is given by
\begin{eqnarray}
 {\cal F} &=& \int_0^L dy \left[\frac{1}{2} \left| \frac{dx(y)}{dy} - h
\right|^2 + U(x(y)) \right]   \\   \label{eq:secondH_f}
    &=& \nonumber -    h \left( x(L) - x(0)  \right) +
\int_0^L dy \left[\frac{1}{2} \left|\frac{dx(y)}{dy} \right|^2 \right.
  \\  &  & \left.  + \, \rule{0in}{16pt} U(x(y)) \right] + \mbox{constant}
\end{eqnarray}
where we have set the overall energy scale $\tilde\epsilon=1$ for
simplicity.  Since we are neglecting thermal effects, the physics is
completely independent of this choice.

This free energy is also the expression for the classical action of a
particle of mass $1$ in 1+1 dimensions in a potential $-U[x]$ where
$y$ corresponds to the time coordinate for the classical particle.
Note that this mapping is somewhat different from the usual boson
analogy.\cite{Seung}\ Since $ -h \left( x(L) - x(0) \right)$ is a
boundary term, it does not effect the equations of motion for the
classical particle.  The quantity
\begin{equation}
\label{eq:EOM}
\tilde E  = \frac{1}{2} \left| \frac{dx}{dy} \right|^2 - U(x(y)).
\end{equation}
corresponds to the energy of the classical particle system and is thus
conserved.  This means  that the local slope  of
the flux line $\frac{dx}{dy}$ is a function only of the local
potential $U[x]$ and of a single constant of the motion $\tilde E $.

The density of states for the flux line potential $U$ is
\begin{equation}
  D(u) = \lim_{W \rightarrow \infty} \frac{1}{W} \int_{-W/2}^{W/2} dx
  \,\,\, \delta(u - U[x])
\end{equation}
For now, we will require that the potential $U$ is bounded from below,
which means there exists a
\begin{equation}
u_{min} = \min U > -\infty.
\end{equation}
Note that $D(u) = 0$ for $u < u_{min}$.

We will compute the tilt response, defined as the slope
\begin{equation}
\theta(h) \equiv \frac{x(L;h) - x(0;h)}{L}
\end{equation}
of the flux line in its ground state.  For small angles, $\theta
\approx \langle B_x\rangle/\langle B_y\rangle$.  If $h=0$, the minimum energy
solution is $x(y) \equiv x_0$ where $ U[x_0] = u_{min}$, i.e. the flux
line in pinned to the deepest point of the random potential.  This
yields a free energy of
\begin{equation}
  \label{eq:Hpinned}
  {\cal F} = u_{min} L
\end{equation}

If the magnetic field is applied at a nonzero slope  $h$, it is
possible that the flux line remains pinned on the deepest point,
yielding the exact same value for the Hamiltonian.  However, it is
also possible that the flux line becomes unpinned and traverses the
system at some average slope  $\theta$ defined as above.  Using
Eq.\ref{eq:EOM}, the magnitude of the local slope  is given by
\begin{equation}
  \left| \frac{dx}{dy} \right| = \sqrt{2
    \left[ \tilde E + U(x(y)) \right]}.
\label{slopemagnitude}
\end{equation}
It is a simple exercise to show that, if the upper endpoint $x(L)$ is
free and the lower endpoint is at an absolute minimum of $U(x)$, the
minimal path with always have $dx/dy \ge 0$ for $h>0$.  It should be
noted that, for an unpinned path ($dx/dy > 0$), $\tilde E > -u_{min}$,
so that the argument of the square root is nonnegative.  In the
language of the classical particle this is just the requirement that
the total energy of the particle is larger than the potential of the
highest hill, so it has enough energy to traverse the system.
Integrating Eq.\ref{slopemagnitude}, the inverse slope $\alpha \equiv
\frac{1}{\theta} = \frac{L}{x(L) - x(0)}$ is
\begin{equation}
  \alpha(\tilde E ) =   \frac{1}{x(L) - x(0)} \int_{x(0)}^{x(L)} dx
\frac{dy}{dx} .
\label{alphaequation}
\end{equation}
In the limit of a large system when the slope $\theta$ is nonzero, the
flux line samples all of the disorder in the potential equally.  Thus
since the probability that a given
point $x$ has potential $U[x] = u$ is just the density of states
$D(u)$, Eqs.\ref{slopemagnitude}\ and \ref{alphaequation} give
\begin{equation}
  \alpha(\tilde E )
  = \int du \frac{D(u)}{\sqrt{2 \left( \tilde E  +
       u \right) }} , \label{eq:alpha(E)}
\end{equation}
so long as $\theta \ne 0$.  This relation implicitly determines
$\tilde E $ as a function $\alpha$.  The slope is
determined  by optimizing over all possible values of $\tilde E$.
It should be noted that  Eq.\ref{eq:alpha(E)} is
completely independent of any spatial correlations in the function
$U(x)$ so long as the system is large enough that the flux line
samples all of the disorder equally.

By using Eq.\ref{eq:EOM}\ to eliminate $U(x)$, Eq.\ref{eq:secondH_f}\
can be rewritten, up to an additive constant, as
\begin{equation}  \label{eq:free2}
  {\cal F} = -h (x(L) - x(0)) - \tilde E L +  \int_{x(0)}^{x(L)} dx
\frac{dx}{dy},
\end{equation}
Taking $L \rightarrow \infty$,
\begin{eqnarray}
  \nonumber {\cal F}(\tilde E ,h) &=& L \left[ -\frac{ h}{\alpha(\tilde E )} -
\tilde E  \right.  \\ &+& \left.
  \frac{1}{\alpha(\tilde E )} \int du \, D(u) \,
  \sqrt{2\left( \tilde E  + u \right)} \right].
  \label{eq:HofE}
\end{eqnarray}
Extremizing ${\cal F}$ with respect to
$\tilde E$ gives
\begin{equation}
  \label{eq:h(E)}
  h =  \int du \, D(u) \,
  \sqrt{2\left( \tilde E  + u \right)}
\end{equation}
which in principle, can be inverted to yield $\tilde E(h)$ and the
tilt response $\theta(h)$ (cf. Eq.\ref{eq:alpha(E)}).  For future
use, we note in passing that
\begin{equation}
  \label{eq:dhdE}
  \frac{dh}{d\tilde E } = \alpha(\tilde E ).
\end{equation}

For some values of $h$, there will be no $\tilde E $ satisfying
Eq.\ref{eq:h(E)}. In this case, there is no finite optimal value of
the inverse slope $\alpha$, and the ground state of the flux line is
pinned with $x(y)=x_0$, and $U[x_0] = u_{min}$ so that $\theta =
1/\alpha = 0$.  As $h$ is increased from zero, the slope $\theta$
remains zero (pinned) until $h$ reaches some critical value $h_c$,
where $\theta$ begins to increase (unpinned) to nonzero values.  We
therefore have $\partial B_x/\partial H_x|_{h=0} =0$, i.e. there is a
transverse Meissner effect.

The behavior near $h_c$ constitutes a critical phenomenon analogous to
that at the lower critical field.  Using the optimal value of $\tilde
E $ found in Eq.\ref{eq:h(E)} in Eq.\ref{eq:HofE}, the optimal free
energy for a given $h$ is
\begin{equation}
 \min {\cal F}(\tilde E ,h) = -  \tilde E (h) L
\end{equation}
as $L\rightarrow\infty$.  This free energy for the flux line is equal
to the energy of a pinned flux line (Eq.\ref{eq:Hpinned}) when
$\tilde E $ obtains its minimal value $-u_{min}$.  In terms of the
language of the classical particle, this simply says that the
classical particle becomes ``unpinned'' as soon as it has sufficient
energy $\tilde E $ to get over the highest hills of the potential.  We
will be able to expand the necessary equations around the critical
value $\tilde E = -u_{min}$ and more closely examine the transitional
behavior.

\subsubsection{An Explicit Example}

To illustrate these results,  consider a disorder potential with the
uniform density of states given by
\begin{equation}
  D(u) = \left\{ \begin{array}{lll}
   -1/u_{min}  & \mbox{for} & u_{min} < u < 0  \\
    0 &   &  \mbox{otherwise}
  \end{array} \right.
\end{equation}
with $u_{min} < 0$.  Eq.\ref{eq:alpha(E)} can be explicitly integrated
to yield
\begin{equation}
  \label{eq:alphee}
  \alpha(\tilde E ) = \frac{\sqrt{2}}{-u_{min}}\left[ \sqrt{\tilde E } -
  \sqrt{\tilde E  + u_{min}} \right].
\end{equation}
As discussed above, the critical value of $\tilde E $ is given by
$\tilde E_c = -u_{min}$, so the critical value of the slope of the
flux line is
\begin{equation}
  \theta_c = \frac{1}{\alpha_c} = \frac{1}{\alpha(\tilde E_c)} =
  \sqrt{\frac{-u_{min}}{2}}.
\end{equation}
Similarly, Eq.\ref{eq:h(E)} can be explicitly integrated to give
\begin{equation}
  \label{eq:star1}
  h(\tilde E ) = \frac{2\sqrt{2}}{-3 u_{min}}
  \left[\tilde E^{3/2} - (\tilde E  + u_{min})^{3/2} \right]
\end{equation}
which yields a  critical slope  of the applied field
\begin{equation}
  h_c = h(\tilde E_c) = \frac{2}{3} \sqrt{-2 u_{min}}.
\end{equation}
For slopes of the applied field that are less than $h_c$, the flux
line is pinned at $\theta =0$.  When $h$ reaches $h_c$, $\theta$ jumps
to $\theta_c$, and then increases continuously as $h$ is further
increased.  In this particular case, Eq.\ref{eq:alphee} can also be
explicitly inverted to yield
\begin{equation}
  \label{eq:starstar1}
  \tilde E (\alpha) = \frac{1}{4} \left[ \frac{\sqrt{2}}{\alpha} -
\frac{\alpha u_{min}}{\sqrt{2}}
\right]^2,
\end{equation}
By substituting Eq.\ref{eq:starstar1} into Eq.\ref{eq:star1}, the
complete (albeit messy) analytic form for the function $h(\alpha)$ can
be found.  In general, however, $\alpha(\tilde E )$ is not explicitly
invertible, so an analytic form for $h(\alpha)$ can not be obtained.
Furthermore, the complete function $h(\alpha)$ is clearly very
sensitive to the precise form of the density of states.  Thus, we
would like to focus on physics that is in some sense more universal.

\subsubsection{Angular Exponent Near The Depinning Transition}

To this end we consider the behavior near the critical slope.  For $h
- h_c$ small but positive, we should have the relation
\begin{equation}
  \label{eq:htheta}
  (h - h_c) \sim (\theta - \theta_c)^\gamma
\end{equation}
for some exponent $\gamma$.  As long as $\theta_c$ is nonzero, this
then implies
\begin{equation}
\label{eq:halpha}
  (h-h_c) \sim (\alpha - \alpha_c)^\gamma.
\end{equation}
In general, the value of the exponent $\gamma$ will only depend on
behavior of the density of states $D(u)$ very near to $u_{min}$.  In the
above example, using the analytic form for $h(\alpha)$, it is trivial
to expand around $\alpha_c$ to get $\gamma = 2$.  In fact, $\gamma$
will take the same value for all systems where the density of states
$D(u)$ has a finite jump discontinuity at $u_{min}$.

Let us suppose that the density of states takes the form
\begin{equation}
\label{eq:Dscale}
  D(u_{min}) \sim (u - u_{min})^\beta
\end{equation}
for $u$ close to but greater than $u_{min}$ for some exponent $\beta$.
Of course we must have $\beta > -1$ such that the integral of $D$ is
finite and $D$ is normalizable.

We are now concerned with the behavior for $\tilde E $ close to but greater
than $-u_{min}$.  We define
\begin{equation}
  \Delta \tilde E= \tilde E  + u_{min}.
\end{equation}
With $\Delta \tilde E$ small, Eq.\ref{eq:alpha(E)} implies (with $p
= u - u_{min}$)
\begin{equation}
  \alpha(\tilde E) \sim \int_0^{q}    dp \, p^\beta (p + \Delta \tilde
E)^{-1/2}
  + \cdots
\end{equation}
where $q$ is some small cutoff beyond which $D(u)$ no longer has the
scaling form \ref{eq:Dscale}.  This integral is finite for $\beta >
-\frac{1}{2}$, and otherwise diverges as
\begin{equation}
  \alpha(\tilde E) \sim (\Delta \tilde E)^{(\beta+\frac{1}{2})}  ~~~  \mbox{for
 $ \beta < -\frac{1}{2}$}.
\end{equation}
On the other hand, Eq.\ref{eq:h(E)} implies that $h(\Delta E)$ is
always finite.  But from Eq.\ref{eq:dhdE}, we have $dh/d\Delta \tilde
E \sim \alpha$ finite as $\Delta \tilde E\rightarrow 0$ only for
$\beta > -\frac{1}{2}$ as described above.  Finally,
\begin{equation}
  \label{eq:alphaint}
  \frac{d \alpha}{d \tilde E} \sim  \int_0^{q}    dp \, p^\beta (p + \Delta
\tilde E)^{-3/2}
  + \cdots
\end{equation}
which is finite for $\beta > \frac{1}{2}$ and diverges as
\begin{equation}
  \frac{d \alpha}{d \tilde E} \sim (\Delta \tilde
  E)^{(\beta-\frac{1}{2})} ~~~ \mbox{for $ \beta < \frac{1}{2}$}.
\end{equation}

In the case of $\beta > \frac{1}{2}$, we have $h$, $\alpha= dh/d\tilde
E$, and $d \alpha/ d \tilde E$ all finite and nonzero as $\Delta
\tilde E \rightarrow 0$.  Thus, expanding around $\Delta \tilde E= 0$,
we have $(h-h_c) \sim (\alpha - \alpha_c) \sim (\theta - \theta_c)
\sim \Delta \tilde E$, so $\gamma = 1$, with $h_c > 0$ and $\theta_c =
1/\alpha_c >0$.

In the case of $\frac{1}{2} > \beta > -\frac{1}{2}$, $dh/d\tilde E =
\alpha$ is again finite as $\Delta \tilde E\rightarrow 0$, so $(h-h_c)
\sim \Delta \tilde E$, but $d\alpha/d \tilde E$ diverges as
$\Delta \tilde E^{(\beta-1/2)}$, so $(\alpha - \alpha_c) \sim \Delta
\tilde E^{(\beta+1/2)}$.  Thus $\gamma = 1/(\beta + \frac{1}{2})$ with
$h_c > 0$ and $\theta_c >0$.  Here we recover the result $\gamma =2 $
for $\beta=0$ that we derived in the explicit example above.

In the case of $-1 < \beta < -\frac{1}{2}$, $\alpha = dh/d\tilde E$
is divergent as $\Delta \tilde E^{(\beta +1/2)}$ as $\Delta \tilde
E\rightarrow 0$, so $(h-h_c) \sim \Delta \tilde E^{(\beta + 3/2)}$.
Here $\alpha_c$ is divergent, so $\theta_c = 0$.  Thus, $\gamma = -
\frac{\beta + 3/2}{\beta+1/2}$, and Eq.\ref{eq:halpha} does not
hold.

For $\beta$ approaching $-\frac{1}{2}$, the exponent $\gamma$
diverges.  For $\beta = -\frac{1}{2}$, Eq.\ref{eq:alphaint} can be
integrated for small $\Delta \tilde E$ to yield $\alpha \sim
-\ln(\Delta \tilde E) + \cdots$ which is divergent for small $\Delta
\tilde E$ so $\theta_c =0$.  Since $dh/d\tilde E = \alpha$, we have
$(h-h_c) \sim \Delta \tilde E (\ln \Delta \tilde E - 1)$.  Thus,
\begin{equation}
  (h-h_c) \sim e^{-1/\theta} \left( \frac{1}{\theta} -1 \right) ~~~
  \mbox{for} ~~~\beta = -\frac{1}{2}
\end{equation}

\subsubsection{Finite Sized Systems}

In the case of a finite sized system, a flux line with nonzero slope
can only sample all of the disorder equally if $|x(L) - x(0)|$ is much
larger than the correlation length of the disorder.  If $|x(L) -
x(0)|$ is not much larger than the correlation length, we can not use
the above method to average over the disorder, and we must treat the
correlations explicitly.

For example, in any finite sized system with smooth disorder (ie,
$U(x)$ has nonzero correlation length), there is some genericly unique
point $x_0$ such that $U(x_0) = u_{min}$.  Near this point, we can
expand $U(x)$ as
\begin{equation}
  U(x) = u_{min} + A (x-x_0)^2 + \ldots.
\end{equation}
For $h=0$, the lowest energy solution is the flux line pinned at
$x_0$. For nonzero $h$, the flux line should pull away from this
minimum in a continuous manner.  In the simple case of $U(x)=A x^2$,
(choosing $u_{min}=0$, and $x_0=0$), we can study this behavior
explicitly.  For simplicity we fix one end of the flux line to be
pinned in the middle of this deepest well, that is, $x(0) = x_0 =0$,
and we allow $x(L)$ to vary.  Then $L$, $\tilde E$, and $x(L)$ are all
related by the restriction
\begin{equation}
  L = \int_{x(0)}^{x(L)} dx \frac{dy}{dx} = \int_{x(0)}^{x(L)} dx
\frac{1}{\sqrt{2(\tilde E + U(x))}}
\end{equation}
which can be integrated explicitly in this parabolic case and solved
to yield
\begin{equation}
 \tilde E = A [x(L)]^2 \left[ \sinh \left(L \sqrt{2A } \right) \right]^{-2}.
\end{equation}
Now differentiating Eq.\ref{eq:free2} with respect to $x(L)$ (and
again using Eq.\ref{slopemagnitude} to find a minimum of the free
energy, yields the general result (compare to Eq.\ref{eq:h(E)})
\begin{equation}
  h = \sqrt{2\left[\tilde E + U(x(L))\right]},
\end{equation}
which in the case of the parabolic potential yields
\begin{equation}
  \label{eq:xres1}
  x(L) = h \left[\sqrt{2A} \coth\left(L \sqrt{2A}
\right)  \right].
\end{equation}
This calculation can equally well be performed in the case where we
allow both endpoints to vary and the result is similar.  Of course
this result holds only in the range where $x(L)$ is small enough such
that the potential $U$ still looks parabolic.  Furthermore, if $x(L)$
becomes sufficiently large that there is a point $x_1$ far away such
that $U(x_1) < U(x(L))$ then it is possible that the flux line will
stretch to a more favorable position near $x_1$.  Nonetheless, for a
generic finite sized system, for $h$ sufficiently small, the slope
$\theta = [x(L)-x(0)]/L$ will vary as $\theta \sim h$.  However, as
can be seen from Eq.\ref{eq:xres1}, as $L$ gets larger, the range of
$h$ values for which this parabolic approximation is valid is reduced
exponentially.  For larger values of $h$ or for larger $L$, the system
may then cross over to behavior more like the infinite sized system
discussed above.

\section{Conclusion}
\label{Conclusion}

In this paper, we have discussed two types of commensurability effects
in Josephson junctions containing columnar defects.  The first kind is
tuned by changing the magnitude of the applied magnetic field along
the direction of the defects.  For periodic arrangements of the pins,
the vortex lattice induced by an applied magnetic field locks in to
reciprocal lattice vectors which are integral multiples of the defect
array wavevector (fractionally commensurate states also occur, but are
suppressed by several powers of $a/\lambda_J$, as discussed in
appendix A).  Near these states, the critical current of a typical
junction is substantially enhanced, though the critical current
density remains zero in the thermodynamic limit.

A second type of commensurability can be investigated by varying the
transverse component of the field, or its direction.  This effect is
present for many distributions of the defects, though the
character of the transition to the incommensurate (unaligned) state
depends upon this distribution.  In the case of periodic defects, the
mapping of section \ref{PeriodicDefects}\ shows that a commensurate
vortex lattice remains locked in until a finite threshold transverse
field is reached, after which kinked fluxons enable the array to tilt.
The transverse constitutive relation $B_\perp(H_\perp)$ was found,
under rather general conditions, to be simply a rescaled version of
the zero parallel field result.  We also investigated the dilute limit
(near $H_{c1J}$) where each vortex acts independently.  Using a
mapping to one--dimensional particle mechanics, we showed that in most
random distributions of pins, the fluxon makes a {\sl first order}
jump to a non--zero tipping angle as the transverse field is increased
to a critical value.  Above this field, this angle increases
continuously, and near the critical field this increase can be
described by an almost--universal scaling exponent.  Similar
discussions of this ``tilt response'' have been given for the case of
a finite density of fluxons with randomly placed defects in
Ref.\onlinecite{HNV}.  In a periodic array of defects away from
commensuration, and in the single vortex limit, the effect does not
exist at finite temperatures, as discussed in Ref.\onlinecite{BN}.

Recent experimental work on long overlap junctions has confirmed many
of our theoretical conclusions for the case of the periodic defect
array.\cite{Itzler,Itzler2}\ In particular, the critical currents at the
commensurate peaks agree quite well with the estimate following
Eq.\ref{lambdaeffectiveitzler}, with small deviations which are still
well understood within the model.  Although the peak widths are less
reliably obtained, they appear to be consistent with
Eqs.\ref{peakwidthsinscreenedlimit}\ and
\ref{peakwidthsinunscreenedlimit}.

A particularly dramatic consequence was reported in
Ref.\onlinecite{Itzler2}.  There, the I--V characteristics were
measured above $I_c$ at various commensurate magnetic fields.  These
I--V curves showed ``commensurate field steps,'' analogous to the zero
field steps seen at zero field.  Upon rescaling the current axis by
$[\tilde{\lambda}_J(n=1)/\tilde{\lambda}_J(n=0)]^2$, the steps for
$n=0$ and $n=1$ collapsed to a single curve. This dynamic measurement
confirms the generality of the mapping.  Another potential
characterization is the number of zero field steps (which is roughly
$L/2\tilde{\lambda}_J$), but an accurate determination requires an
impractically large system width.  It would be interesting to further
explore experimentally and theoretically the consequences of this
mapping.

Another interesting possibility from the theoretical point of view is
a ``floating'' phase, in which a commensurate fluxon array is
prevented from locking into the periodic defect array by thermal
fluctuations.  As discussed in \ref{JJreview}, the large value of
$\epsilon_J/(k_{\rm B}T)$ renders thermal fluctuations unimportant
unless $1-T/T_c \ll 1$.  Such a floating phase may, however, be
possible near the bulk superconducting transition, along with other
interesting fluctuation effects not discussed here.  Experimental
observation of such a state would no doubt be difficult due precisely
to these noisy fluctuations.  We must leave these experimental and
theoretical questions open. \\

\centerline{\bf ACKNOWLEDGMENTS}
\vspace{10pt}

It is a pleasure to acknowledge helpful conversations with Mark
Itzler, Michael Tinkham, and Onuttom Narayan.  We would also like to
thank Mark Itzler for making his experimental data available to us
prior to publication.  This research was supported by the National
Science Foundation under Grant No. PHY89--04035 at the ITP, and
National Science Foundation Grant DMR-91-15491 at Harvard.

\appendix
\section{Half--Integer Commensurate States}

In this appendix, we extend the treatment of section
\ref{PeriodicDefects}\ to handle the behavior near the half--integer
commensurate fields.  This condition is defined (at the center of a
peak) by
\begin{equation}
H_y^{(k/2)} = {k \over 2}{{q\phi_0} \over {2\pi d}},
\label{halfintpeaks}
\end{equation}
where $k$ is an odd integer (for $k$ even this condition reduces to
that for the integral peaks discussed earlier).  Higher denominator
fractional commensurate states should also occur, but the treatment
becomes considerably more complicated and will not be described here.

Proceeding as before, the change of variables $\gamma = px + \eta$,
with $p=kq/2$, leads to the equation
\begin{eqnarray}
\lambda_J^2\nabla^2\eta & = & \sum_{m=0}^\infty {\alpha_m \over 2} \bigg\{
\sin[(p+mq)x + \eta] \nonumber \\
& & + \sin[(p-mq)x + \eta] \bigg\},
\label{etaequation}
\end{eqnarray}

Unlike Eq.\ref{expandedSG}, Eq.\ref{etaequation}\ does not contain any
non--oscillatory terms.  More precisely, a first order perturbative
expansion in $\alpha$ will give convergent results.  Nevertheless, we
expect that the fluxon array should lock in to the defect lattice for
small $\tilde{H}_y$.  In fact, the half--integer commensurate effect
shows up as a divergence in the {\sl second order} perturbative
solution of Eq.\ref{etaequation}.\cite{BSunpublished}\ To understand
the nature of this divergence, we make the (non--linear) change of
variables from $\eta$ to $\sigma$, where
\begin{equation}
\eta = \eta_0(\sigma,x) + \sigma,
\label{halfintcovs}
\end{equation}
where
\begin{eqnarray}
\eta_0 & = & -\sum_m {\alpha_m \over {2\lambda_J^2}}
\bigg[{{\sin[(p+mq)x+\sigma]}
\over {(p+mq)^2}} \nonumber \\
& & + {{\sin[(p-mq)x+\sigma]} \over {(p-mq)^2}}\bigg].
\label{eta0def}
\end{eqnarray}
To motivate Eq.\ref{halfintcovs}, note that $\eta_0$ is the
lowest order solution which would have been obtained from
Eq.\ref{etaequation}, taking a constant value of $\eta=\sigma$ on the
right hand side.  Despite the non--linearity,
\begin{eqnarray}
\nabla^2\eta & \approx & \nabla^2\sigma + \sum_m {\alpha_m \over 2}
\bigg[ \sin[(p+mq)x+\sigma] \nonumber \\
& & + \sin[(p-mq)x + \sigma] \bigg],
\label{approxlaplacian}
\end{eqnarray}
up to oscillatory terms which may be neglected to lowest non--trivial
order in $1/(\lambda_J p)$ for small $\tilde{H}_y$.  At this level of
approximation, Eq.\ref{etaequation} becomes
\begin{eqnarray}
\lambda_J^2\nabla^2\sigma & & = \sum_m {\alpha_m \over 2} \bigg[
\sin[(p+mq)x + \eta_0 + \sigma] \nonumber \\
& & + \sin[(p-mq)x + \eta_0 + \sigma] - \sin[(p+mq)x + \sigma]
\nonumber \\
& & - \sin[(p-mq)x + \sigma] \bigg].
\label{sigmaequation1}
\end{eqnarray}
The leading (linear) contribution in a Taylor series in $\eta_0$ gives
the result
\begin{eqnarray}
\lambda_J^2\nabla^2\sigma & = & \sum_{m=0}^\infty {\alpha_m \over 2}
\eta_0\bigg[\cos[(p+mq)x + \sigma] \nonumber \\
& & + \cos[(p-mq)x + \sigma] \bigg].
\label{sigmaequation2}
\end{eqnarray}
By inserting Eq.\ref{eta0def}\ into Eq.\ref{sigmaequation2},
combining sines and cosines, and neglecting oscillatory terms
independent of $\sigma$, one finds
\begin{eqnarray}
\lambda_J^2\nabla^2\sigma & = & \sum_{m,m'} {{\alpha_m\alpha_{m'}} \over
{8\lambda_J^2}} \bigg\{ {1 \over (p-m'q)^2}
\bigg[\sin\Big([(m+m')q-2p]x-2\sigma\Big) - \sin\Big([(m-m')q+2p]x
+ 2\sigma\Big)\bigg] \nonumber \\
&& - {1 \over (p+m'q)^2}\bigg[\sin\Big([(m+m')q+2p]x+2\sigma\Big)-
\sin\Big([(m-m')q-2p]-2\sigma\Big)\bigg]\bigg\}.
\label{awfulequ}
\end{eqnarray}
Keeping only the unmodulated contributions proportional to
$\sin(2\sigma)$, we arrive at the effective equation
\begin{equation}
\tilde{\lambda}_J^2\nabla^2\tilde{\sigma} = \sin\tilde{\sigma},
\label{halfintequation}
\end{equation}
where $\tilde{\sigma} \equiv 2\sigma$ and the effective
Josephson penetration depth in this case is
\begin{eqnarray}
\tilde{\lambda}_J & = & \lambda_J \bigg\{
{{\alpha_k\alpha_0} \over {4\lambda_J^2 p^2}} - {1 \over
{4\lambda_J^2}} \sum_{m=0}^\infty \bigg[
{{\alpha_m\alpha_{|m-k|}} \over {|p-(k-m)q|^2}} \nonumber \\
& & + {{\alpha_m\alpha_{m+k}} \over {|p-(k+m)q|^2}}\bigg] \bigg\}^{-1/2} .
\label{halfinteffectivelambdaj}
\end{eqnarray}
It should be noted that the change of variables (Eq.\ref{halfintcovs})
from $\eta$ to $\sigma$ induces a complicated transformation of the
boundary conditions in Eq.\ref{bcs}.  The corrections are, however,
small by a factor of $1/(\lambda_J p)^2$.  Although we have not
performed a detailed analysis of their effects upon the JJ, we
therefore expect only minor quantitative changes in behavior.
Physically, one effect of these corrections is to account for overlap
with the long
``Fraunhofer'' tails of the integer--field (including zero--field)
commensurate peaks.  A crude way of accounting for this effect is
to treat the critical current, calculated as in
Eq.\ref{commensuratecritcurr}, as a shift relative to the
(experimentally determined) Fraunhofer background.

Neglecting these complications, and taking into account the final
change of variables from $\sigma$ to $\tilde{\sigma}$, the effective
current and effective fields are
\begin{eqnarray}
\tilde{H}_y & = & 2(H_y - H_y^{(k/2)}), \nonumber \\
\tilde{H}_x & = & 2H_x, \nonumber \\
\tilde{I} & = & 2I,
\label{effectiveparametershalfint}
\end{eqnarray}
where $H_y^{(k/2)}$ is the commensurate field defined in
Eq.\ref{halfintpeaks}.

{}From Eq.\ref{halfinteffectivelambdaj}, it is clear that the
half--integer states have critical currents suppressed by a factor of
$(a/\lambda_J)^2$.  We expect that higher denominator fractional
states can also occur, but are suppressed by even higher powers of
$a/\lambda_J$.

We have estimated the magnitude of the critical current peak in the
experiments of Ref.\onlinecite{Itzler}.  Using Eqs.\ref{itzlercase},
\ref{halfinteffectivelambdaj}, \ref{commensuratecritcurr}, and
\ref{effectiveparametershalfint}, we obtain a lowest 1/2 integral peak
reduced by a factor of approximately $600$ from the zero--field
critical current.  Using $I_{c0} = 30 mA$ as determined
experimentally, this gives $\Delta I_c \approx 0.05 mA$.  A feature is
barely visible in the critical current versus field plot (Fig.2 of
Ref.\onlinecite{Itzler}), with a deviation from the background of
roughly $0.2 mA$.  Given the crude treatment of the boundary
conditions and the smallness of the effect, such order of magnitude
agreement is satisfactory.

\begin{figure}
\epsfxsize=3.5truein
\hskip 0.0truein \epsffile{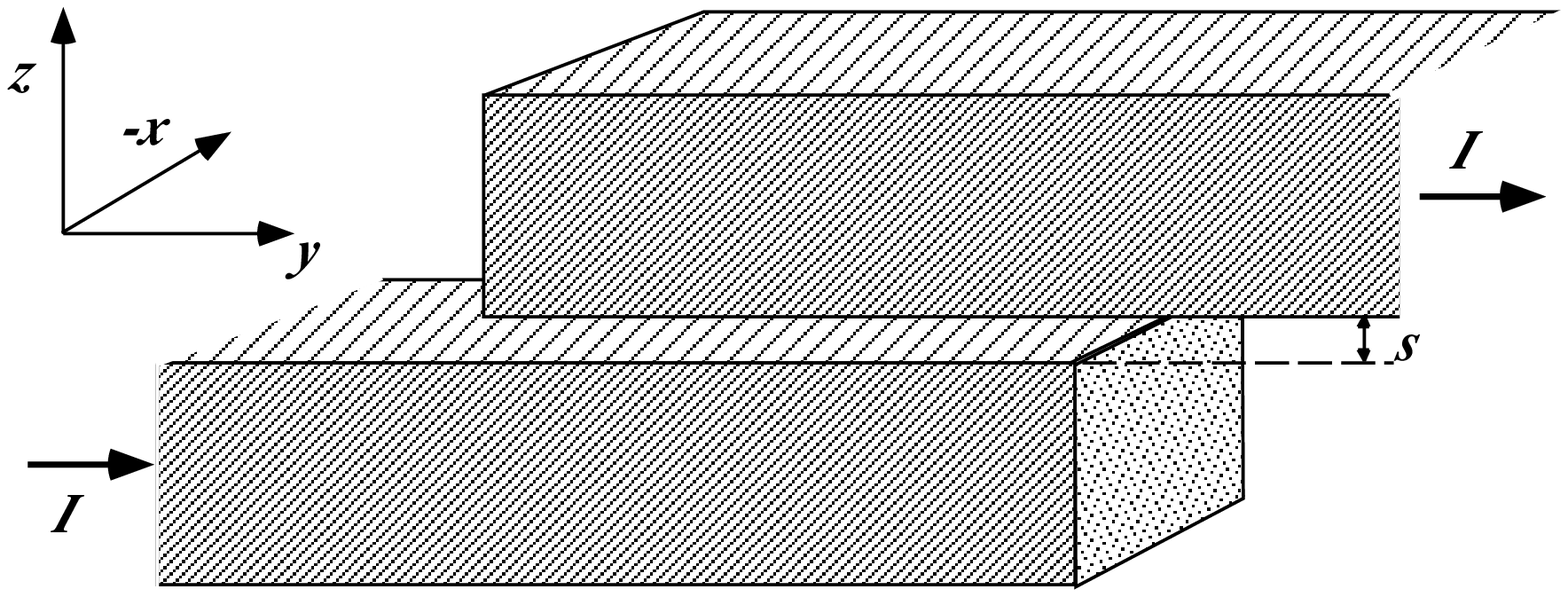}
 { FIG 1: Overlap geometry for a large Josephson junction.}
\label{overlapfig}
\end{figure}


\begin{references}

\bibitem{VGtheory} D. S. Fisher, M. P. A. Fisher, and D. A. Huse,
Phys. Rev. {\bf B43}, 130 (1991).

\bibitem{BGtheory} D. R. Nelson and V. M. Vinokur, Phys. Rev.
B{\bf 48}, 13060 (1993).

\bibitem{VGexperiments} R. H. Koch, V. Foglietti, W. J. Gallagher, G.
Koren, A. Gupta, and M. P. A. Fisher, Phys. Rev. Lett. {\bf 63}, 1511
(1989); P. L. Gammel, L. F. Schneemener, and D. J. Bishop, Phys. Rev.
Lett. {\bf 66}, 953 (1991).

\bibitem{BGexperiments} L. Civale, A. D. Marwick, T. K. Worthington,
M. A. Kirk, J. R. Thompson, L. Krusin-Elbaum, Y. Sum, J. R. Clem and
F. Holtzberg, Phys. Rev.  Lett. {\bf 67}, 648 (1991); M. Leghissa,
L. A. Gurevich, M. Kraus, G.  Saemann--Ischenko, and L. Ya. Vinnikov,
Phys. Rev. B {\bf 48}, 1341 (1993).


\bibitem{numerics} Recent numerical work may yield better estimates of
vortex and Bose glass critical exponents.  See A. P. Young, J. Phys. A
{\bf 26}, L1067 (1993); H. S. Bokil and A. P. Young, unpublished.

\bibitem{RGrefs} T. Natterman, I. Lyuksyutov, and M. Schwartz,
Europhys. Lett. {\bf 16}, 295 (1991); J. Toner, Phys. Rev. Lett. {\bf
67}, 2537 (1991); Y.-C. Tsai and Y. Shapir, Phys. Rev. Lett. {\bf 69},
1773 (1992).

\bibitem{HwaBatrouni} G. G. Batrouni and T. Hwa, Phys. Rev. Lett. {\bf
72}, 4133 (1994).

\bibitem{FGB} R. Fehrenbacher, V. B. Geshkenbein, and G. Blatter,
Phys. Rev. B{\bf 45}, 5450 (1992).

\bibitem{VK} V. M. Vinokur and A. E. Koshelev, Zh. Eksp. Teor. Fiz.
{\bf 97}, 976 (1990) [Sov. Phys. JETP {\bf 70}, 547 (1990)].

\bibitem{Itzler} M. A. Itzler and M. Tinkham, Phys. Rev. B., in press.

\bibitem{Itzler2} M. A. Itzler and M. Tinkham, to appear in {\em IEEE
Trans. Appl. Supercon.}, (1994).  See also M. A. Itzler, unpublished.

\bibitem{Seung}  D. R. Nelson, Phys. Rev. Lett. {\bf 60}, 1973 (1988);
D. R. Nelson and H. S. Seung, Phys. Rev. B {\bf 39}, 9153 (1989).

\bibitem{Tinkham} See, e.g. M. Tinkham, {\it Introduction to
Superconductivity,}  (McGraw-Hill, New York, 1975).

\bibitem{OS} C. N. Owen and D. J. Scalapino, Phys. Rev. {\bf 164}, 538
  (1967).

\bibitem{JJgeneral} See, e.g. A. Barone, ed., {\it Josephson Effect:
Achievements and Trends} (world Scientific Pub. Co., Singapore, 1986).

\bibitem{lineenergynote} The asymptotic form (including prefactors) of
$\tilde{\epsilon}$ and $V(\chi)$ is obtained by substituting
$\gamma(x,y) = \sum_i\gamma_0(x-x_i(y))$ in Eq.\ref{freeenergy}, where
$\gamma_0(x)$ is given in Eq.\ref{exactonesoliton}.

\bibitem{CIT} See, e.g. S. N. Coppersmith, D. S. Fisher, B. I.
Halperin, P. A. Lee, and W. F. Brinkman, Phys. Rev. B {\bf 25}, 349
(1982).


\bibitem{thicknessnote} The thickness dependence can be incorporated
by including an additional term $-\lambda_J^2 d^{-1} \nabla d
\cdot \nabla\gamma$ on the left hand side of Eq.\ref{modifiedSG}.



\bibitem{pathology} This applies only to configurations of defects in
which pinning  sites do not become arbitrarily closely spaced as the
junction size is increased.  If this restriction is relaxed, as in the
``optimized'' junction of Ref.\onlinecite{FGB}, a finite critical
current density is indeed possible.

\bibitem{longnote} For a long JJ of the overlap geometry considered in
most of this paper, however, $\bar{j}$ need not vanish for $W
\rightarrow \infty$.  This is because the system is actually two
dimensional, but not fully screened in the $y$ direction.  If $L$ is
taken to infinity as well, $\bar{j}$ will indeed vanish.

\bibitem{integralstatenote} This transformation is appropriate for
studying the behavior near the ``integral'' commensurate states with
$H_y = {{nq\phi_0} \over {2\pi d}}$.  ``Fractional'' states can also
occur, and are discussed in appendix A.

\bibitem{correctionsnote} In fact, these terms give non-singular
  perturbative contributions, which are small in the limit considered.
  It is easy to show that the corrections are of order
  $(q\lambda_J)^2$.

\bibitem{HNV} T. Hwa, D. R. Nelson, and V. M. Vinokur, Phys. Rev. B
{\bf 48}, 1167 (1993).

\bibitem{BN} L. Balents and D. R. Nelson, submitted to Phys. Rev.
Lett., (1994).

\bibitem{BSunpublished} L. Balents and S. H. Simon, unpublished.

\end{references}
\end{document}